\documentclass[12pt,thmsa]{article}

\usepackage{amsfonts}
\usepackage{graphicx}
\usepackage{epsfig}
\usepackage{graphics}

\makeatletter
\newcommand{\row}[1]%
{\mathord{\buildrel{\lower3pt%
\hbox{$\scriptscriptstyle\rightarrow$}}\over #1}}

\newcommand{\dyadic}[1]{\mathord{\dyadic@rrow{#1}}}
\newcommand{\dyadic@rrow}[1]{
\begin{picture}(12,12)(-1,0)
\put(-1,9){\makebox(0,0)[t]{$\scriptscriptstyle\downarrow$}}
\put(-1,9){\makebox(0,0)[l]{$\scriptscriptstyle\longrightarrow$}}
\put(5,0){\makebox(0,0)[b]{$#1$}}
\end{picture}
}

\newcommand{\bra}[1]{\bigl\langle #1 \bigr|}
\newcommand{\ket}[1]{\bigl| #1 \bigr\rangle}

\begin{document}
\begin{center}

{\large Sudden death and rebirth of Entanglement for Different
Dimensional Systems driven by a Classical Random External Field}
\\
N. Metwally$^a$, H.Eleuch$^b$, A.-S. Obada$^c$\\
$^{a}$Mathematics Department, College of Science, Bahrain
University,
Bahrain\\
$^{a}$Mathematics Department, College of Science, Aswan
University, Aswan, Egypt
\\
$^{b}$Department of Physics, McGill University, Montreal, Canada
H3A 2T8
\\
$^{c}$Mathematics Department, Faculty of Science, Al-Azhar
University, Cairo, Egypt
\\
email:nmetwally@gmail.com
 \vspace{10pt}
\end{center}

\begin{abstract} The entangled behavior of different dimensional
systems driven by classical external random field is investigated.
The amount of the survival entanglement between the components of
each system is quantified. There are different behaviors of
entanglement  that come into view  decay, sudden death, sudden
birth and long-lived entanglement. The maximum entangled states
which can be generated from any of theses suggested systems are
much fragile than the partially entangled ones. The  systems of
larger dimensions are more robust than  those of  smaller
dimensions systems, where the entanglement decay smoothly,
gradually and  may vanish for a very short time.  For the class of
$2\times 3$ dimensional system, the one parameter family is found
to be more robust than the two parameters family. Although the
entanglement of driven $ 2 \times 3$ dimensional system is very
sensitive to the classical external random field, one can use them
to generate a long-lived entanglement.
\end{abstract}

pacs:03.67.-a, 03.67.Bg, 03.67.Hk
 \vspace{2pc}
 \noindent{\it keywords}:External fields, Qubit, qutrit, Entanglement


\section{Introduction}

It is well known that, noise represents one of the unavoidable
physical phenomena in the context of quantum information and
computation. The effect of different types of noises on many
systems has been investigated extensively for two dimensional
systems, (see for example
\cite{Ban,Eberly2006,Metwally2009,Jia2014,Aguilar2012,Aguilar2014}). The
dynamics of higher dimensional systems which travel in different
noise channels  have been discussed. For example, the separability
of entangled qutrits in a  noise channel is investigated in
\cite{Agata2007}. The time evolution of classical and quantum
correlations of hybrid qubit-qutrit systems in a classical
dephasing environment  has been studied \cite{Karpat2011}. Also,
the decoherent dynamics of quantum correlations in qubit-qutrit
system under various  noise channels has been discussed in
\cite{Liang2013}. The dynamics of entanglement for a qutrit system
in the presence of global, collective, multilocal and local noise
channels is studied \cite{Khan}. The possibility of protecting
entanglement in qubit-qutrit system from decoherence via weak
measurements and reversal was considered by Xiao \cite{Xiao}. The
phenomenon of bipartite entanglement revivals under local
operations in systems subject to classical noise sources is
investigated  \cite{Franco0,Franco01}.

Entangled systems which are driven by classical field usually lose
their correlations and consequently their efficiency to perform
quantum communication decreases. Recently this type of study has
implemented experimentally \cite{Franco00,Franco2}. Most of these
studies  focused on small dimensional systems (qubits). For
example, ref \cite{Metwally2012}  investigated the dynamics of the
encoded information in a pulsed-driven qubit within and outside the
rotating wave approximation. The dynamics of the tripartite
entanglement, where two qubits are driven non-resonantly coupled
to the cavity is discussed \cite{Blanter2012}. The effect of
rectangular and exponential pulses on the degree of correlations
between two qubit systems was discussed in \cite{Metwally2014-1}.

Larger dimensional systems  as $2\times 3$ and $3\times 3$, which
are  driven by classical  external random fields  have not been
investigated in details. There are only few research works in this
direction that have been carried out. Recently, it has been shown
that, an external classical driving field  for a qubit can  speed-up the evolution of an open system. \cite{Ying}.Therefore, we are motivated to investigate the entanglement degradation between different  or similar
dimensional subsystems which are driven by classical external
random field (CERF). This study is devoted for qubit-qubit,
qubit-qutrit and qutrit-qutrit systems, where it is assumed that,
only one particle is driven by CERF. The main aim of this paper is
 quantify the  survival entanglement between the
subsystems  of each system when only one subsystem is driven
by CERF.

This paper is organized as follows: In Sec.$2$, the suggested
models and their evolution are introduced. The analytical
solutions are given in Sec.$3$ as well as the behaviors of
entanglement for different initial states settings. We summarize
our results in Sec.$4$.

\section{Models}
The suggested model consists of three  different systems,
two-qubits system, which represents a $2\times 2$ dimensional system,
qubit-qutrit ($2\times 3$) system which consists of two and three
levels subsystems and two-qutrit system, where each subsystem
 is defined by $3\times 3 $ dimensions. Fig.(1) represents
 schematic diagram for the suggested models.

\subsection{Dynamics of 2-qubit systems}

\begin{figure}[t!]
 \centering
    \includegraphics[width=30pc,height=20pc]{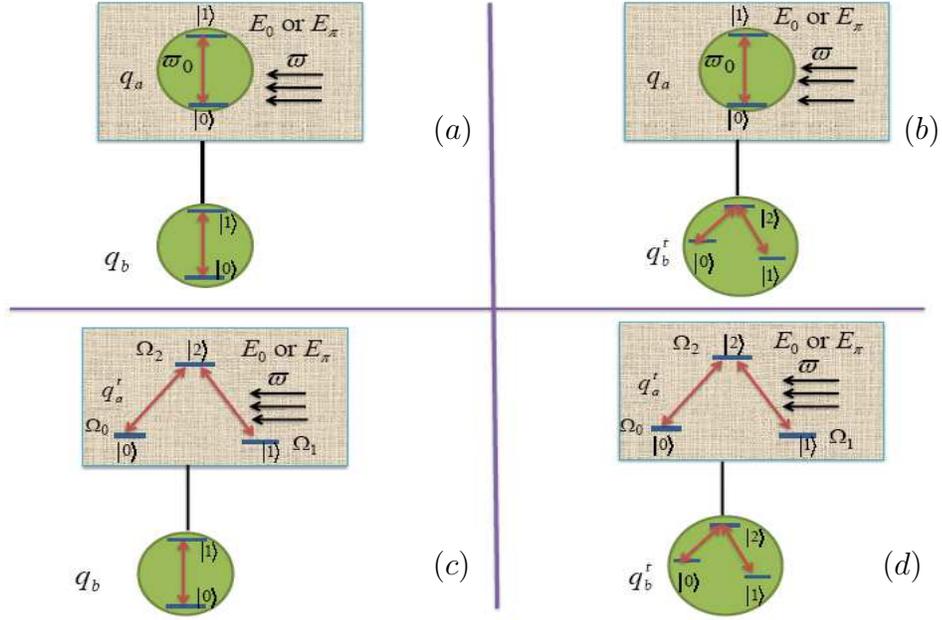}
     \put(-190,185){$(a)$}
      \put(-12,185){$(b)$}
      \put(-190,20){$(c)$}
      \put(-20,20){$(d)$}
                  \caption{Schematic diagram of the suggested models. (a) A system consists of two qubits one of them is
          driven by classical external random field with frequency $\omega$.
          (b) A qubit-qutrit system where the qubit is driven by CERF with frequency
          $\omega$ and the qutrit is prepared in $\Lambda$ configuration with one upper state
          $\ket{2}$ with energy $\Omega_2$
           and two lowers $\ket{0}$ and $\ket{1}$ with energies $\Omega_0$ and $\Omega_1$, respectively.
           (c) A qubit-qutrit system,  where the qutrit is driven by the CERF with frequency $\omega$
           (d) A system consists of two-qutrits
            prepared in $\Lambda$ configuration defined as (c), where only one qutrit is driven by CERF. }
  \end{figure}

      It is assumed that, one of the subsystems of each  system interacts  locally     with  its own environment which is  described by  a single classical random external field (CERF). This field has a frequency $\omega$ and a random phase $\phi$ which equals either $0$ or $\pi$ with
probability $P=0.5$. The schematic description is shown in
Fig.(1a). In the rotating frame approximations, the Hamiltonian,
$\mathcal{H}$ of the single qubit-field system is given by,
\begin{equation}
\mathcal{H}_{q-f}=\frac{\omega_0-\omega}{2}\sigma_z+i\hbar
g(\sigma_+e^{-i\phi}-\sigma_{-}e^{i\phi}),
\end{equation}
where $g$ is  the coupling strength between the qubit and the
CREF, $\sigma_{\pm}$ are the raising and lowering operators of the
single qubit with frequency $\omega_0$.  For the suggested
2-qubit systems, it is assumed that,
\begin{itemize}
 \item Only one qubit is driven
by the CERF. \item Only the resonances case is considered i.e.,  $\omega_0=\omega$.
\item  The interaction between the qubit
and its environment is strong enough, where the dissipation
between the vacuum and the qubit is forbidden.

\end{itemize}
By considering the above assumptions, the final state at $t>0$ can
be written as,
\begin{equation}\label{H2D}
\rho_{2\times
2}(t)=\frac{1}{2}\sum_{j=1}^{2}\left\{e^{-i\mathcal{H}_jt}\rho_{2\times
2}(0)e^{i\mathcal{H}_jt}\right\},
\end{equation}
where $\mathcal{H}_j=i\hbar g_i(\sigma_{+}e^{-i\phi_j}-
\sigma_{-}e^{i\phi_j})$, j=1,2 with  $\phi_1=0$ and $\phi_2=\pi$,
respectively and we set $\hbar=1$ for simplicity. The initial
state $\rho_{2\times 2}(0)$ represents the  state of  a system
consists of $2\times 2$ dimensions, i.e., 2-qubits system.

 \subsection{ Dynamics  of qubit-qutrit systems}
 This model  represents a $2\times 3$ dimensional  system. It is
 assumed  that, only one  subsystem is driven by the CERF. However,
 if the qubit is driven by the classical external random field, then the dynamics of the system
 is governed  by  Eq.(\ref{H2D}) (see Fig.(1b)). On the other hand, if we allow  the $3D$
 subsystem, namely the qutrit, to be driven by the   local CERF,
 then the Hamiltonian which describes this system depends on the
 configuration of the qutrit-system (see Fig.(1c)). Let us consider
 that,  the qutrit is initially prepared in $\Lambda$ configuration \cite{Obada0}. In this case, the
 Hamiltonian which governs the interaction between the CERF and the single qutrit is given by,
 \begin{eqnarray}
 \mathcal{H}_{qt-f}&=&\sum_{\ell=0}^{2}\omega_{\ell}
 S_{\ell}+g_1(e^{-i\phi}S_{12}+e^{i\phi }S_{21})
+ g_2(e^{-i\phi }S_{20}+e^{i\phi }S_{02}),
\nonumber\\
&&=\mathcal{H}_0+\mathcal{H}_{int},
 \end{eqnarray}
where $\omega_{\ell}, g_{\ell}, \ell=1,2$ are the frequencies of
the external fields and the coupling strength between the field
and the qutrit. The operators $S_{ij}$, where $(i,j)\in {12,21,20,02}$, obey the
$SU(3)$ algebra. For this suggested system, we consider the
following assumptions:
 \begin{itemize}
  \item The  single qutrit system has one upper level
 $\ket{2}$ with frequency $\Omega_2$ and two lower levels $\ket{0}$ and $\ket{1}$
 with frequencies $\Omega_0$ and $\Omega_1$, respectively.
 \item The transitions between the
 $\ket{0}\leftrightarrow\ket{2}$ and
 $\ket{1}\leftrightarrow\ket{2}$ are dipole allowed, while
 between  $\ket{0}\leftrightarrow\ket{1}$ is dipole forbidden,
 i.e., there is no interaction.
 \item The non-degeneracy case is considered \cite{Marlan,Aguilar}, namely
 $(\Omega_2-\Omega_1)-(\Omega_2-\Omega_0)\gg g_\ell$, $\ell=1,2$.

\item The resonance case is considered i.e.
$\Omega_2-\Omega_0=\omega_1$, $\Omega_2-\Omega_1=\omega_2$.
\end{itemize}
By considering the above assumptions, the dynamics of the
qubit-qutrit  system  is given by,
\begin{eqnarray}
\rho_{2\times 3}(t)&=&
\frac{1}{2}\sum_{i=1}^{2}\left\{e^{-i\mathcal{H}_{int}t}\rho_{2\times
3}(0)e^{i\mathcal{H}_{int}t}\right\},\quad\mbox{with}
\nonumber\\
\mathcal{H}_{int}&=&g_1(e^{-i\phi_i}S_{12}+e^{i\phi_i }S_{21})+
g_2(e^{-i\phi_i }S_{20}+e^{i\phi_i }S_{02}), ~i=0, \pi.
\end{eqnarray}
\subsection{Dynamics of qutrit-qutrit system}
In  this case, the initial system consists of two
 $-$three levels subsystems, namely two qutrits, which represent a
 generalization of a qubit. The state of the qutrit can be spanned
 by the three orthogonal basis $\ket{0}, \ket{1},\ket{2}$.
 Physically, qutrit can be represented by three-level atoms
 \cite{Obada1,Obada2,Obada3} (see Fig.(1d)).  In this treatment, we
 consider only one qutrit is driven by the CERF, where
 both qutrits are initially  prepared in $\Lambda$ configuration.
 Moreover, we consider the same assumptions  listed
 above(qubit-qutrit case). The dynamics  of the qutrit-qutrit
 system $\rho_{3\times 3}(0)$ is given by,
 \begin{equation}
 \rho_{3\times 3}(t)=\frac{1}{2}\sum_{i=1}^{2}\left\{e^{-i\mathcal{H}_{int}t}\rho_{3\times3}(0)e^{i\mathcal{H}_{int}t}\right\},
\end{equation}
 where $\mathcal{H}_{int}$ is given from Eq.(4).

\section{Survival Entanglement}
In this section, we obtain   analytical solutions for all  the
above systems. The behavior of entanglement between each two
subsystems  will be  discussed, where  the amount of entanglement
is quantified  by using a  measure called negativity. This
measure is an acceptable measure for any bipartite system of any
dimension \cite{Lee}. For a state $\rho_{12}$, with dimensions
$d_1\times d_2$ where $d_1<d_2$, the negativity, $\mathcal{N}$ is
defined as,
\begin{equation}
\mathcal{N}=\frac{1}{d_1-1}\left\{||\rho_{12}^{T_2}||-1\right\},
\end{equation}
where $\rho_{12}^{T_2}$ is the partial transpose with respect to
the largest party (second-partite) and $||.||$ is the trace norm \cite{Lee,Karpat}

\subsection{Qubit-qubit systems}

Let us assume that, the two  qubits state is given by $X$-state,
which
 can be described by the computational  basis  $"0"$ and $"1"$
as,
\begin{eqnarray}
\rho_x(0)&=&\frac{1+c_{33}}{2}\Bigl(\ket{00}\bra{00}+\ket{11}\bra{11}\Bigr)
+\frac{1-c_{33}}{2}\Bigl(\ket{01}\bra{01}+\ket{10}\bra{10}\Bigr)
\nonumber\\
&&+
\frac{c_{11}+c_{22}}{2}\Bigl(\ket{01}\bra{10}+\ket{10}\bra{01}\Bigr)
+
\frac{c_{11}-c_{22}}{2}\Bigl(\ket{00}\bra{11}+\ket{11}\bra{00}\Bigr),
\end{eqnarray}
where $c_{ij}=tr\{\rho_x(0)\sigma_i\sigma_j\}, i,j=1,2,3$ and
$\sigma_{i,j}$ are the Pauli matrices for the first and the second
qubits, respectively. From this state, one can obtain a maximum
entangled classes of states  by setting $c_{ij}=\pm 1$, Werner
state by setting $c_{ij}=x$ and a generalized Werner state for
$c_{11}=x_1, c_{22}=x_2$ and $c_{33}=x_3$.

\begin{figure}
  \begin{center}\label{qubit}
   \includegraphics[width=25pc,height=15pc]{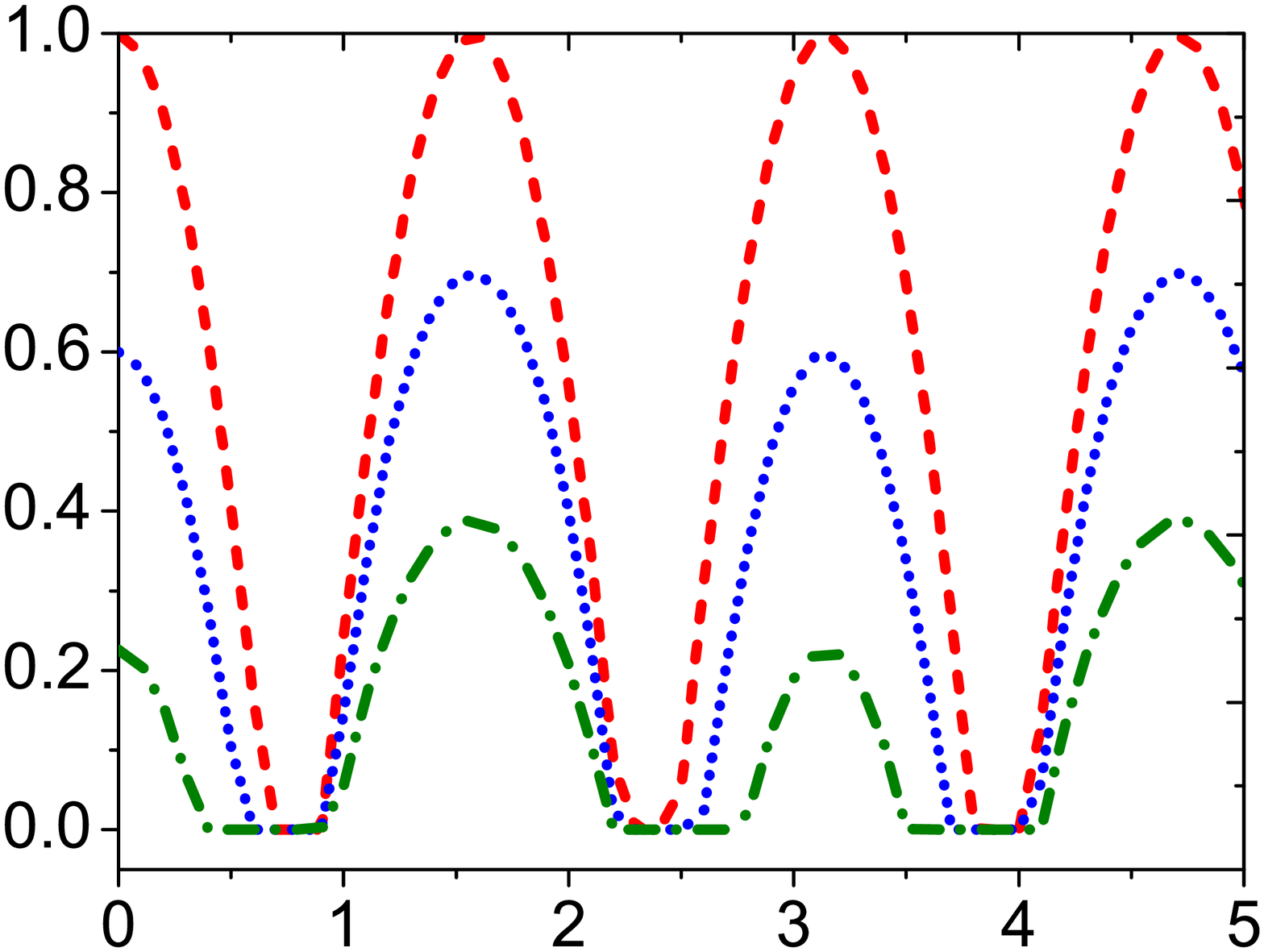}
 \put(-145,5){$\tau$}
     \put(-275,90){$\mathcal{N}$}
      \caption{The evolution of the   entanglement for two dimensional system (two-qubits) against $\tau$. The
      dash, dot and dash-dot  curves for MES,( $c_{11}=c_{22}=c_{33}=-1$), Werner state with $x=-0.8$
      and a generalized Werner state ( $c_{11}=-0.8, c_{22}=-0.7$, and $c_{33}=-0.6$), respectively .}
  \end{center}
\end{figure}
If only one qubit is driven by the classical external random
field, then the final state of the system at  a time $\tau>0$ is
given by,
\begin{eqnarray}
\rho_x(t)&=&\mathcal{R}_1\ket{00}\bra{00}+\mathcal{R}_2\ket{00}\bra{10}+\mathcal{R}_3\ket{00}\bra{11}
+\mathcal{R}_4\ket{01}\bra{01}+
\mathcal{R}_5\ket{01}\bra{10}+\mathcal{R}_6\ket{10}\bra{01}
\nonumber\\
&&+\mathcal{R}_7\ket{10}\bra{10}+\mathcal{R}_8\ket{10}\bra{11}+\mathcal{R}_9\ket{11}\bra{00}
+\mathcal{R}_{10}\ket{11}\bra{10}+\mathcal{R}_{11}\ket{11}\bra{11},
\end{eqnarray}
where,
 \begin{eqnarray}
\mathcal{R}_1&=&\frac{1+c_{33}}{2}cos^2(\tau)+\frac{1-c_{33}}{2}sin^2(\tau),
\nonumber\\
 \mathcal{R}_2&=&-\frac{1+c_{33}}{2}\sin(\tau)\cos(\tau),
\nonumber\\
\mathcal{R}_3&=&\frac{c_{11}-c_{22}}{2}cos^2(\tau)-\frac{c_{11}+c_{22}}{2}sin^2(\tau),
\nonumber\\
\mathcal{R}_4&=&\frac{1-c_{33}}{2}cos^2(\tau)+\frac{1+c_{33}}{2}sin^2(\tau),
\nonumber\\
\mathcal{R}_5&=&\frac{c_{11}+c_{22}}{2}cos^2(\tau)-\frac{c_{11}-c_{22}}{2}sin^2(\tau)=\mathcal{R}_6,
\nonumber\\
\mathcal{R}_7&=&\frac{1-c_{33}}{2}cos^2(\tau)-\frac{1+c_{33}}{2}sin^2(\tau)
\nonumber\\
\mathcal{R}_8&=&-\frac{c_{11}+c_{22}}{4}\sin(2\tau),\quad\mathcal{R}_9=\mathcal{R}_3,
\nonumber\\ \quad
\mathcal{R}_{10}&=&-\frac{c_{11}-c_{22}}{4}\sin(2\tau),\quad
\mathcal{R}_{11}=\mathcal{R}_1.
\end{eqnarray}
where $gt=\tau$

In Fig.(2), we investigate the effect of the classical external
random  field on   three different classes of two-qubit systems:
maximum entangled class (MES), where we set
$c_{11}=c_{22}=c_{33}=-1$, Werner state, with
$(c_{11}=c_{22}=c_{33}=x=-0.8)$ and partially entangled classes
(PES) with $c_{11}=-0.8,c_{22}=-0.7,c_{33}=-0.6$. The general
behavior shows that, the entanglement decays as  the time
increases for all these initial states. The phenomena of sudden
death of entanglement is depicted for each MES and PES. However,
the time death increases for initially less entangled states. For
lager interaction time, the entanglement re-birth again to reach
its maximum bounds which depend on the entanglement of the initial
states. For partially entangled states, the upper bounds of
entanglement are larger than the initial entanglement in some
intervals of time.  Moreover, the upper bounds are reached at the
same time for all states.

\subsection{Qubit-Qutrit systems}
In this subsection,a  system consisting  of two different
dimensional subsystems is considered: one is a  qubit (2D) and the
other is a qutrit (3D). Analytical solutions for the final states
are obtained for different families. The first state  is known by one
parameter family \cite{Karpat} and the second  known by a two
parameters  family \cite{Chi}. The time evolution is obtained when
only one  of their subsystems is driven by classical external
random field.

\subsubsection{  One parameter family\\}

 This state represents a
qubit-qutrit system which is  defined by $2\times 3$ dimensions.
It is  described by one parameter as,
\begin{eqnarray}\label{1p}
\rho_{1p}(0)&=&\frac{\mathcal{P}}{2}\bigl(\ket{00}\bra{00}+\ket{01}\bra{01}+\ket{11}\bra{11}\bigr)+
\frac{\mathcal{P}}{2}\bigl(\ket{12}\bra{12}+\ket{12}\bra{00}+\ket{00}\bra{12}\bigr).
\nonumber\\
&&+\frac{1-2\mathcal{P}}{2}\left(\ket{02}\bra{02}+\ket{10}\bra{02}+\ket{10}\bra{10}+\ket{02}\bra{10}\right)
\end{eqnarray}
This state has quantum correlation for $\mathcal{P}\in [0,\frac{1}{3})\bigcup(\frac{1}{3},0.5]$, i.e., it is   disentangled  at $\mathcal{P}=\frac{1}{3}$ only. Now we consider
the following two cases:
\begin{itemize}
    \item {\it Only the qubit is driven by the CERF\\}
    The time evolution of the initial state (\ref{1p}) can be obtained by
    using Eq.(2). For $\tau>0$, the final state $\rho_{1p}(t)$ can be written  explicitly
    as,
    \begin{eqnarray}
\rho_{1p}^{q_b}(t)&=&\mathcal{L}_1^{q_b}\ket{00}\bra{00}+\mathcal{L}_2^{q_b}\ket{00}\bra{12}+\mathcal{L}_3^{q_b}\ket{01}\bra{01}
+\mathcal{L}_4^{q_b}\ket{02}\bra{02}+\mathcal{L}_5^{q_b}\ket{10}\bra{10}
\nonumber\\
&&+\mathcal{L}_6^{q_b}\ket{10}\bra{02}+\mathcal{L}_7^{q_b}\ket{02}\bra{10}
+\mathcal{L}_8^{q_b}\ket{11}\bra{11}+\mathcal{L}_9^{q_b}\ket{12}\bra{12}
+\mathcal{L}_{10}^{q_b}\ket{12}\bra{00}, \nonumber\\
\end{eqnarray}
where, the superscript $q_b$ refers to the qubit  while
the subscript $1p$  means one parameter family. The coefficients
$\mathcal{L}^{q_b}_j, j=1...10 $ are given by,
\begin{eqnarray}
\mathcal{L}_{1,2}^{q_b}&=&\frac{\mathcal{P}}{2}\cos^2(\tau)\pm\frac{1-2\mathcal{P}}{2}\sin^2(\tau),
\quad \mathcal{L}_3^{q_b}=\frac{\mathcal{P}}{2},
\nonumber\\
\mathcal{L}_4^{q_b}&=&\mathcal{L}_5^{q_b}=\frac{\mathcal{P}}{2}\sin^2(\tau)+\frac{1-2\mathcal{P}}{2}\cos^2(\tau),
\nonumber\\
\mathcal{L}_6^{q_b}&=&\mathcal{L}_7^{q_u}=\frac{1-2\mathcal{P}}{2}\cos^2(\tau)-\frac{\mathcal{P}}{2}\sin^2(\tau),
\nonumber\\
 \mathcal{L}_8^{q_b}&=&\mathcal{L}_3^{q_b}, \quad
\mathcal{L}_9^{q_b}=\mathcal{L}_1^{q_b},\quad
\mathcal{L}_{10}^{q_b}=\mathcal{L}_1^{q_b},\mbox{where}~\quad
gt=\tau.
\end{eqnarray}

    \item {\it Only the qutrit is driven by CERF\\}
    In this case, it is assumed that only the qutrit is allowed to  be driven by
    the classical external  random field, where the non-degenerate state is considered \cite{Marlan,Aguilar}.
     The  final state is given by,
   \begin{eqnarray}
   \rho_{1p}^{q_t}&=&\mathcal{L}_1^{q_t}\ket{00}\bra{00}+\mathcal{L}_2^{q_t}\ket{00}\bra{02}+\mathcal{L}_3^{q_t}\ket{00}\bra{10}
  +\mathcal{L}_4^{q_t}\ket{00}\bra{12}+\mathcal{L}_5^{q_t}\ket{01}\bra{00}
\nonumber\\
&&+\mathcal{L}_6^{q_t}\ket{01}\bra{01}+\mathcal{L}_7^{q_t}\ket{01}\bra{10}+\mathcal{L}_8^{q_t}\ket{01}\bra{11}+\mathcal{L}_9^{q_u}\ket{02}\bra{00}
+\mathcal{L}_{10}^{q_t}\ket{02}\bra{02}
\nonumber\\
&&+\mathcal{L}_{11}^{q_t}\ket{10}\bra{00}+\mathcal{L}_{12}^{q_t}\ket{10}\bra{02}+\mathcal{L}_{13}^{q_t}\ket{10}\bra{10}+\mathcal{L}_{14}^{q_t}\ket{10}\bra{11}+\mathcal{L}_{15}^{q_t}\ket{10}\bra{12}
\nonumber\\
&&+\mathcal{L}_{16}^{q_t}\ket{11}\bra{00}+\mathcal{L}_{17}^{q_t}\ket{11}\bra{11}+\mathcal{L}_{18}^{q_t}\ket{11}\bra{10}
+\mathcal{L}_{19}^{q_t}\ket{12}\bra{10}+\mathcal{L}_{20}^{q_t}\ket{12}\bra{12}
\nonumber\\
&&+\mathcal{L}_{21}^{q_t}\ket{00}\bra{11}
+\mathcal{L}_{22}^{q_t}\ket{02}\bra{10}+\mathcal{L}_{23}^{q_t}\ket{10}\bra{01}+
\mathcal{L}_{24}^{q_t}\ket{11}\bra{01}+\mathcal{L}_{25}^{q_t}\ket{12}\bra{00}
\nonumber\\
\end{eqnarray}
where, the subscript $q_t$  refers  the qutrit
The coefficients $\mathcal{L}_j^{q_t}, j=1...25$ are given by,
\begin{eqnarray}
\mathcal{L}_1^{q_t}&=&\frac{\mathcal{P}}{2}\cos^2(\tau_1)+\frac{1-2\mathcal{P}}{2}\sin^2(\tau_1),\quad
\mathcal{L}_2^{q_t}=-\frac{1-2\mathcal{P}}{4}\sin(\tau_1)\sin(2\tau_2),
\nonumber\\
\mathcal{L}_3^{q_t}&=&-\frac{1-\mathcal{P}}{4}\sin(2\tau_1)\sin(\tau_2),\quad
\mathcal{L}_4=\frac{\mathcal{P}}{2}\cos(\tau_1)\cos(\tau_2),\quad
\nonumber\\
\mathcal{L}_5^{q_t}&=&\frac{1-2\mathcal{P}}{8}\sin(2\tau_1)\sin(2\tau_2),
\nonumber\\
\mathcal{L}_6^{q_t}&=&\frac{\mathcal{P}}{2}\Bigl(\sin^2(\tau_1)+\cos^2(\tau_2)\cos^2(\tau_2)\Bigr)
+ \frac{1-2\mathcal{P}}{2}\cos^2(\tau_1)\sin^2(\tau_2),
\nonumber\\
\mathcal{L}_7^{q_t}&=&\frac{\mathcal{P}}{2}\sin^2(\tau_1)\cos(\tau_2),
\quad
\mathcal{L}_8^{q_t}=\frac{1-\mathcal{P}}{4}\sin(2\tau_1)\sin(\tau_2),\quad
\nonumber\\
\mathcal{L}_9^{q_t}&=&-\frac{1-2\mathcal{P}}{4}\sin(\tau_1)\sin(2\tau_2),\nonumber\\
\mathcal{L}_{10}^{q_t}&=&\frac{\mathcal{P}}{2}\sin^2(\tau_2)+\frac{1-2\mathcal{P}}{2}\cos^2(\tau_2),\quad
 \mathcal{L}_{11}^{q_t}=\frac{\mathcal{P}}{2}\sin(\tau_1)\sin(\tau_2)\Bigl(\cos(\tau_1)+\cos(\tau_2)\Bigr),\nonumber\\
\mathcal{L}_{12}^{q_t}&=&\frac{1-2\mathcal{P}}{2}\cos(\tau_1)\cos(\tau_2),\quad
\mathcal{L}_{13}^{q_t}=\frac{\mathcal{P}}{2}\sin^2(\tau_1)+\frac{1-2\mathcal{P}}{2}\cos^2(\tau_2),\nonumber\\
\mathcal{L}_{14}^{q_t}&=&\frac{\mathcal{P}}{8}\sin(2\tau_1)\sin(2\tau_2),\quad
\mathcal{L}_{15}^{q_t}=-\frac{\mathcal{P}}{4}\sin(2\tau_1)\cos(\tau_2),\quad
\nonumber\\
\mathcal{L}_{16}^{q_t}&=&\frac{1-2\mathcal{P}}{2}\sin^2(\tau_1)\cos(\tau_2),\nonumber\\
\mathcal{L}_{17}^{q_t}&=&\frac{\mathcal{P}}{2}\cos^2(\tau_1)+\frac{1-2\mathcal{P}}{2}\sin^2(\tau_2),
\quad \mathcal{L}_{18}^{q_t}=\mathcal{L}_{14}^{q_t}, \quad
\mathcal{L}_{19}^{q_t}=\mathcal{L}_{15}^{q_u},\quad
\mathcal{L}_{20}^{q_t}=\frac{\mathcal{P}}{2},
\nonumber\\
\mathcal{L}_{21}^{q_t}&=&\mathcal{L}_{16}^{q_t},\quad
\mathcal{L}_{22}^{q_t}=\mathcal{L}_{12}^{q_t}, \quad
\mathcal{L}_{23}^{q_t}=\mathcal{L}_{7}^{q_t},\quad
\mathcal{L}_{24}^{q_t}=\frac{\mathcal{P}}{4}\sin(\tau_1)\sin(\tau_2),\quad
\mathcal{L}_{25}^{q_t}=\mathcal{L}_{4}^{q_t}.
\end{eqnarray}
where $g_1t=\tau_1$ and $g_2t=\tau_2$.
\begin{figure}[t!]
  \centering
    \includegraphics[width=20pc,height=15pc]{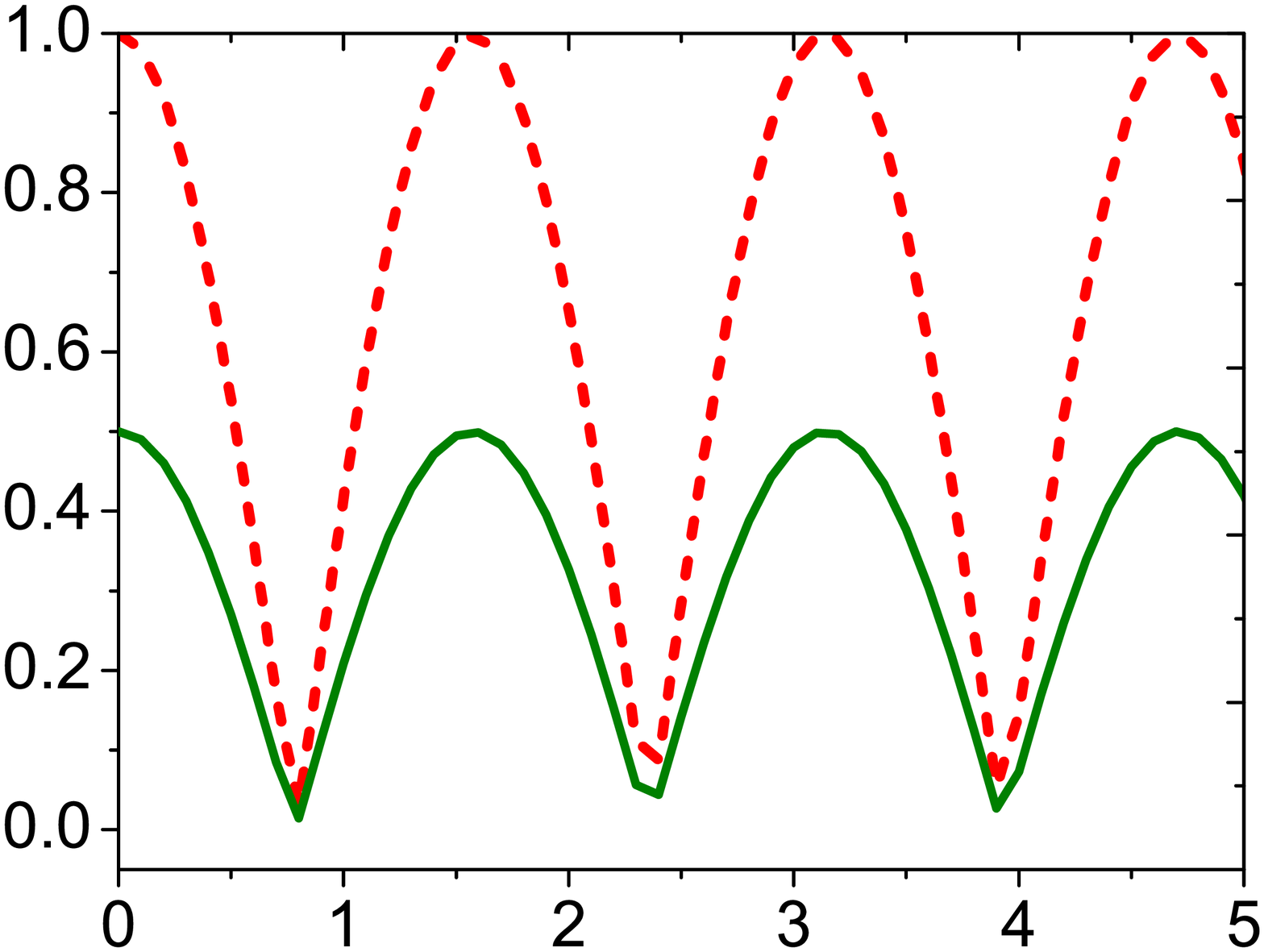}
          \put(-220,90){$\mathcal{N}$}
  \put(-120,5){$\tau'$}
  \put(-70,145){$(a)$}
        \includegraphics[width=20pc,height=15pc]{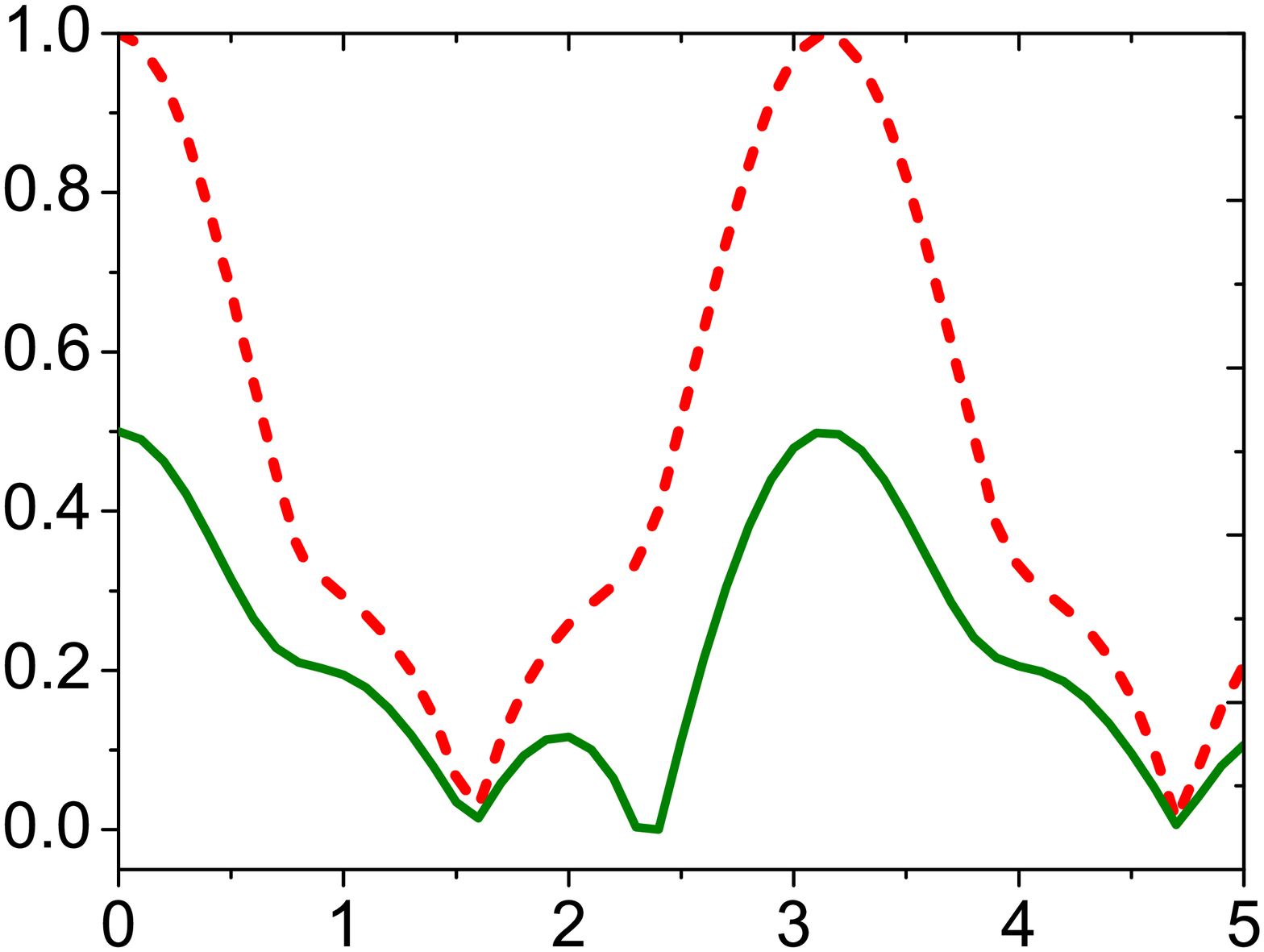}
         \put(-220,90){$\mathcal{N}$}
  \put(-120,5){$\tau'$}
    \put(-50,145){$(b)$}
       \caption{Entanglement of  one parameter family qubit-qutrit system,
       where the dash and dash-dot curves stand  for $p=0$ (MES)
      and $p=0.5$ (PES). (a) qubit is driven (b) qutrit is driven, where we set $\tau_1=\tau_2=\tau'$}
\end{figure}

The behavior of entanglement for a  system is  prepared initially
in a qubit-qutrit  state of one parameter  family type is depicted
in Fig.(3), where it is assumed that only the qubit or the qutrit
are driven by the classical external random field. Two values of
the parameter $\mathcal{P}$ are considered,
$\mathcal{P}=0$ for a system  is initially prepared in a maximum
entangled state \cite{Xing} and $\mathcal{P}=0.5$ for  a
system is initially prepared in a partially entangled state (PES).

Fig.(3a), describes the entanglement behavior between the two
subsystems (qubit and qutrit) when only the qubit is driven by the
CERF. The general behavior shows that, at $\tau=0$, the entanglement
between the two subsystems is maximum, ($\mathcal{N}=1)$. However,
as soon as the interaction is switched on between the qubit and
the CERF, the entanglement decays  fast to reach its minimum bound
$(\mathcal{N}=0.0292)$ for the first time at $\tau=0.8$. As $\tau$
increases further, the entanglement increases suddenly to reach
its maximum value $(\mathcal{N}=1)$ for the first time at $\tau=1.6$.
For larger values of $\tau$, the  behavior of entanglement is
repeated periodically, where the entanglement between the two
subsystems never vanishes. On the other hand, if we start from a
partially entangled state ($\mathcal{P}=0.5$), the behavior of
entanglement is similar to that predicted for MES. However, the
lower bounds of entanglement are larger than those depicted for the two qubit system (MES), while the death time is smaller.

Fig.(3b) shows the dynamics of entanglement when only the qutrit
is driven by the CERF, where  the same initial state settings are
considered, ($\mathcal{P}=0$ and $0.5$). The general behavior
shows that, the entanglement decreases as soon as  the interaction is
switched on. The decay rate is much smaller than that shown in
Fig.(3a), where only the qubit is driven by the CERF. It is clear
that, the entanglement decays slowly and the phenomena of the sudden
changes of entanglement appear clearly \cite{Metwally2014}. The
decay time is much larger than that depicted in Fig.(3a), where
the two subsystems are completely separable for the first time at
$\tau\simeq 1.5$.

\end{itemize}

\subsubsection{Two parameters Family\\}  In  computational basis, the state which describes this family
can be written as \cite{Chi},
\begin{eqnarray}\label{2p}
\rho_{2p}(0)&=&\alpha(\ket{02}\bra{02}+\ket{12}\bra{12})
+\beta(\ket{00}\bra{00}+\ket{11}\bra{11})
+\frac{\beta+\gamma}{2}(\ket{01}\bra{01}+\ket{10}\bra{10})
\nonumber\\
&&+\frac{\beta-\gamma}{2}(\ket{01}\bra{10}+\ket{10}\bra{01}),
\end{eqnarray}
where $\gamma+2\alpha+3\beta=1$. The two
parameters family state (\ref{2p}) is entangled for $\beta=0$,
$\alpha\in[0,\frac{1}{2}]$ and $\gamma\in[\frac{1}{2},1]$.
Moreover, if we set $\alpha=\frac{1}{2}$ and $\gamma=1$, then the
 state (\ref{2p}) turns into a maximum entangled state (MES). Now,
 let us consider the following two cases:

\begin{itemize}
\item{\it Only the qubit is driven by the CERF\\}In this case, the
time evolution of the initial state (\ref{2p}) is given by the
following density operator,
\begin{eqnarray}
\rho_{2p}(t)&=&\mathcal{M}_1\ket{00}\bra{00}+\mathcal{M}_2\ket{01}\bra{01}+\mathcal{M}_3\ket{10}\bra{10}
+\mathcal{M}_4\ket{01}\bra{10}+\mathcal{M}_5\ket{10}\bra{01}
\nonumber\\
&+&\mathcal{M}_6\ket{11}\bra{11}+ \mathcal{M}_7\ket{00}\bra{11}+
\mathcal{M}_8\ket{11}\bra{00}+\mathcal{M}_9\ket{02}\bra{02}
+\mathcal{M}_{10}\ket{12}\bra{12},\nonumber\\
\end{eqnarray}
where,
\begin{eqnarray}
\mathcal{M}_1&=&\beta\cos^2(\tau)+\frac{\beta+\gamma}{2}\sin^2(gt),
\mathcal{M}_2=\beta\sin^2(\tau)+\frac{\beta+\gamma}{2}\cos^2(\tau)
\nonumber\\
\mathcal{M}_3&=&\mathcal{M}_2, \quad
\mathcal{M}_4=\mathcal{M}_5=\frac{\beta-\gamma}{2}\cos^2(\tau),\quad
\mathcal{M}_6=\mathcal{M}_1,\nonumber\\
\mathcal{M}_7&=&\mathcal{M}_8=-\frac{\beta-\gamma}{2}\sin^2(\tau),\quad
\mathcal{M}_9=\mathcal{M}_{10}=\alpha.
\end{eqnarray}
\begin{figure}[t!]
\centering
      \includegraphics[width=20pc,height=15pc]{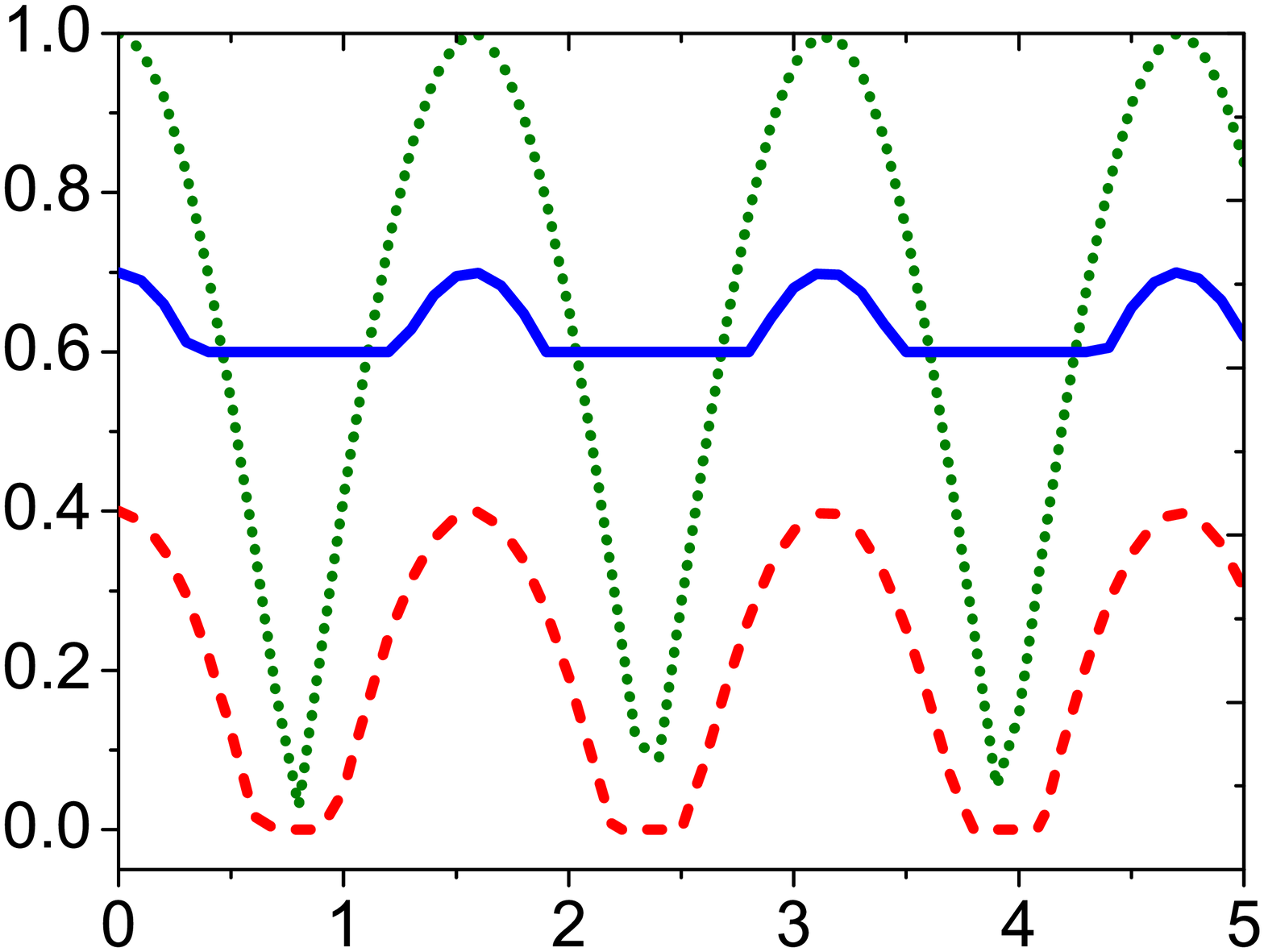}
      \put(-220,90){$\mathcal{N}$}
   \put(-120,5){$\tau'$}
     \put(-70,145){$(a)$}
  \includegraphics[width=20pc,height=15pc]{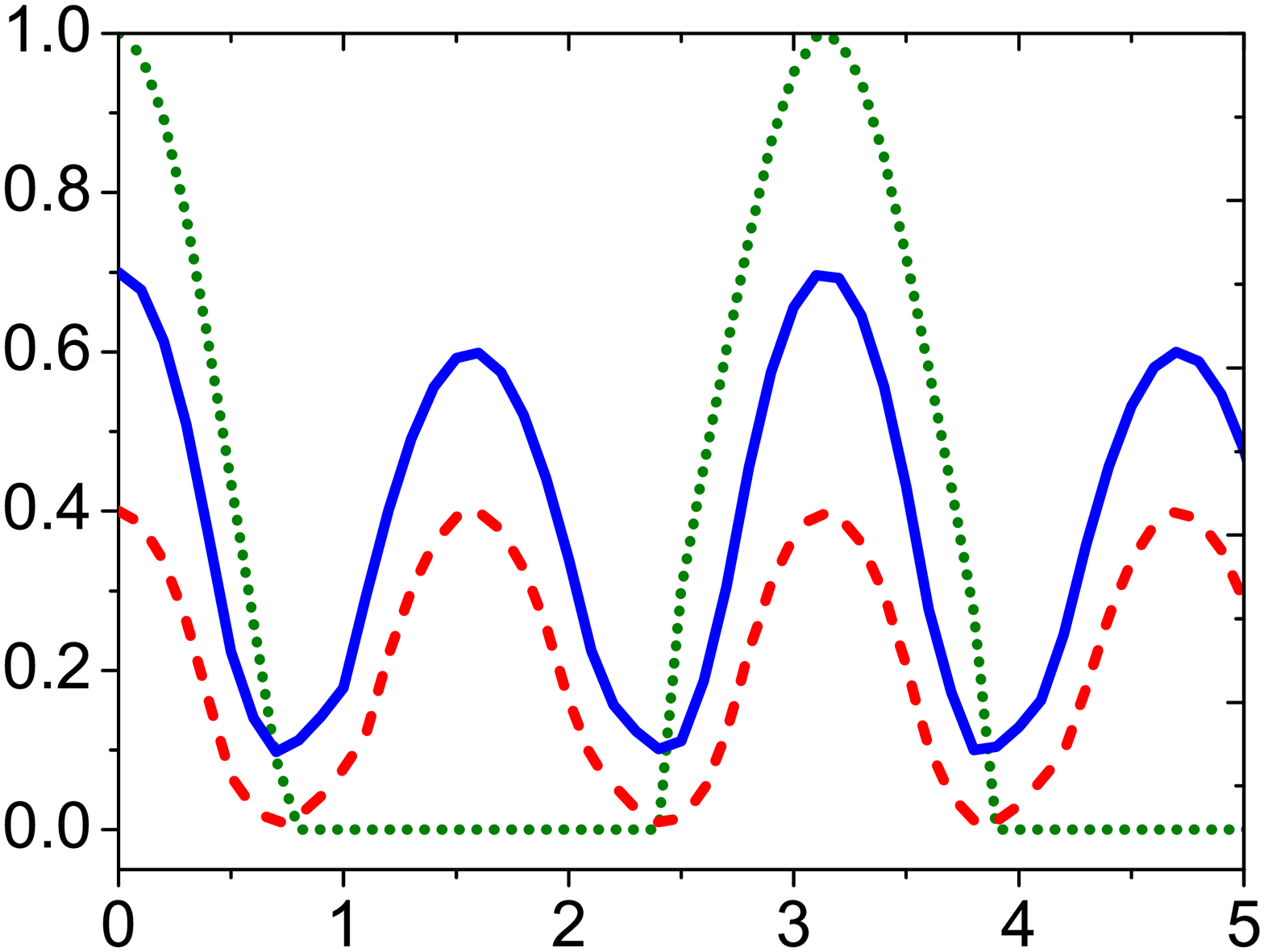}
                \put(-120,5){$\tau'$}
     \put(-220,90){$\mathcal{N}$}
      \put(-50,145){$(b)$}
       \caption{Entanglement of  the two parameters family qubit-qutrit system (a) when only the qubit
       is driven by  CERF and (b) Only the qutrit is driven by the CERF, where $\tau_1=\tau_2=\tau'$. The dot  curve for
       MES with $(\beta=0,\gamma=1,\alpha=0.5)$, while the dash
      and the solid curves for  PES with $(\beta=0.2,\gamma=0.7)$,
        and  $(\beta=0.2,\gamma=0.6)$, respectively.}
 \end{figure}

\item{\it Only the qutrit is driven by the  CERF}
\end{itemize}
In this case, the final state of the system is described by a
matrix of $6\time 6$ dimensions, where  the non-zero elements are
given by,
\begin{eqnarray}
\varrho_{00,00}&=&\beta c_1^2+\alpha s_1^2,\quad
\varrho_{00,01}=\varrho_{01,00}=\frac{\alpha}{4}s_1s_2,\quad
\varrho_{00,02}=-\frac{\alpha}{2}s_1\sin(2\tau_2),\nonumber\\
\varrho_{01,01}&=&\beta s_1^2+c_1^2\left(\alpha
s_2^2+\frac{\beta+\gamma}{2}c_2^2\right), \quad
\varrho_{01,10}=\varrho_{10,01}=\frac{\beta-\gamma}{2}c_1^2c_2,
 \nonumber\\
 \varrho_{02,00}&=& \varrho_{0,02}, \quad
  \varrho_{02,02}=\alpha c_2^2+\frac{\beta+\gamma}{2}s_2^2,
\quad
   \varrho_{02,10}=\frac{\beta-\gamma}{2}s_1s_2
 \nonumber\\
\varrho_{10,10}&=&\alpha s_1^2+\frac{\beta+\gamma}{2}c_1^2, \quad
\varrho_{10,11}=\frac{\alpha}{4}\sin(2\tau_1)\sin(\tau_2),\quad
\varrho_{10,12}=-\frac{\alpha}{2}s_1\sin(2\tau_2)\nonumber\\
 \varrho_{11,02}&=& \varrho_{02,10}, \quad
 \varrho_{11,10}=\varrho_{10,11}, \quad
  \varrho_{12,10}= \varrho_{10,12}, \quad
   \varrho_{11,02}=\alpha c_2^2+s_2^2\beta,
\end{eqnarray}
with $s_i=\sin(\tau_i), c_i=\cos(\tau_i), \tau_i=g_it, ~i=1,2$

In Fig.(4), the amount of survival entanglement between the two
subsystems (qubit and qutrit) is displayed for different initial
state settings. The behavior of entanglement when only the qubit
is driven by  the classical external random field is depicted in
Fig.(4a). If the initial system is initially prepared in a maximum
entangled state, then the entanglement  suddenly decays to reach
its minimum value for the first time at $\tau\simeq 0.8$, then
suddenly increases to be maximum $(\mathcal{N}=1)$. This behavior
is periodically  repeated  as $\tau$ increases. The phenomenon of
long-lived entanglement is predicted for systems that are
initially prepared in partially entangled states, while the
phenomena of sudden decay and death of entanglement  are shown for
systems  that are  initially prepared with small
 entanglement.

Fig.(4b), shows the behavior of entanglement when only the qutrit
is driven by CERF.  In this case, for initially MES, the
entanglement decays very fast to death completely at $\tau'\simeq 0.7$. For this class of states, the phenomenon of sudden birth  of entanglement is depicted, where the death time is
much larger than that shown for one  parameter family. For
initially less entangled state, the entanglement decreases then
increases gradually. The phenomenon  of entanglement death can be
seen only for a short time for systems which have  small initial
 entanglement.

From this figure, one concludes that, systems which are initially
prepared in MES are more  sensitive  to the classical external
random  field than those which  are   prepared in PES.
If the   larger dimensional systems are driven by  the CERF, then the
entanglement of the MES  are liable to  sudden death for a longer
time. Systems which are initially prepared in PES are more robust
than those    prepared in MES for the CERF. The
long-lived entanglement can be observed if one has the ability to
control the parameters which describe the initial  state.
Consequently, one can perform quantum information tasks even in
the presences of these types of noises.

\subsection{Qutrit-qutrit system}
A  system of qutrit-qutrit can be defined by,
\begin{eqnarray}
\rho_{qt}(0)&=&\ket{\psi_{qt}(0)}\bra{\psi_{qt}(0)},\quad
\mbox{where}
 \nonumber\\ &&
 \ket{\psi_{qt}(0)}=a_1\ket{00}+a_2\ket{11}+a_3\ket{22},\quad
\end{eqnarray}
where $a_i,i=1,2,3$ are real and $a_1^2+a_2^2+a_3^2=1$. In this
treatment, it is assumed that both qutrits are prepared in
$\Lambda$ configuration and only the first qutrit is driven by the
classical external random field CERF. By using Eq.(5), one gets
the final state of the qutrit-qutrit system at any value of $\tau_1=\tau_2>0$.
This  final state is defined by a  $9\times 9$ matrix, where the non-zero
elements are given by,
\begin{eqnarray}
\rho_{00,00}&=&a_1^2c_1^2,\quad \rho_{00,11}=a_1a_2c_1^2c_2,\quad
\rho_{00,22}=a_1a_3c_1c_2,\quad
\rho_{00,02}=-a_1a_3c_1s_1s_2,\nonumber\\
\rho_{11,00}&=&a_1a_2c_1^2c_2,\quad
 \rho_{11,11}=a_2^2c_1^2c_2^2,
\quad \rho_{11,22}=a_2a_3c_1c_2^2, \quad
\rho_{00,02}=-a_2a_3c_1c_2s_1s_2,\nonumber\\
\rho_{22,00}&=&a_3a_1c_1c_2, \quad
\rho_{22,11}=a_2a_3c_1c_2^2,\quad
 \rho_{22,22}=a_3^2c_2^2, \quad
\rho_{22,02}=-a_3^2c_2s_1s_2,
\nonumber\\
\rho_{02,00}&=&-a_3a_1s_1s_2c_1,\quad
\rho_{02,11}=-a_3a_2s_1s_2c_1c_2, \quad
\rho_{02,22}=-a_3^2s_1s_2c_2, \quad \rho_{02,02}=a_3^2s_1^2s_2^2
\nonumber\\
\rho_{21,10}&=&a_2a_1s_1s_2,\quad \rho_{21,21}=a_2^2s_2^2,\quad
\rho_{21,12}=a_2a_3s_1s_2c_1, \quad
\rho_{21,02}=a_2a_3s_1^2s_2c_2, \quad \nonumber\\
\rho_{12,10}&=&a_3a_1c_1s_1s_2,\quad
\rho_{12,21}=a_2a_3c_1s_2^2,\quad \rho_{12,12}=a_3^2c_1^2s_2^2,
\quad \rho_{12,02}=a_3^2c_1c_2s_1s_2,
\nonumber\\
\rho_{10,10}&=&a_1^2s_1^2,\quad \rho_{10,21}=a_1a_2c_1s_2,\quad
\rho_{10,12}=a_1a_3^2s_1c_1s_2, \quad \rho_{10,02}=a_1a_3s_1^2c_2
,\quad
\nonumber\\
\rho_{02,10}&=&a_3a_1s_1^2c_2, \rho_{02,21}=a_3a_2s_1s_2c_2,\quad
\rho_{02,12}=a_3^2s_1c_1s_2c_2,\quad \rho_{02,02}=a_3^2s_1^2c_2^2.
\end{eqnarray}

 \begin{figure}
\centering
      \includegraphics[width=25pc,height=15pc]{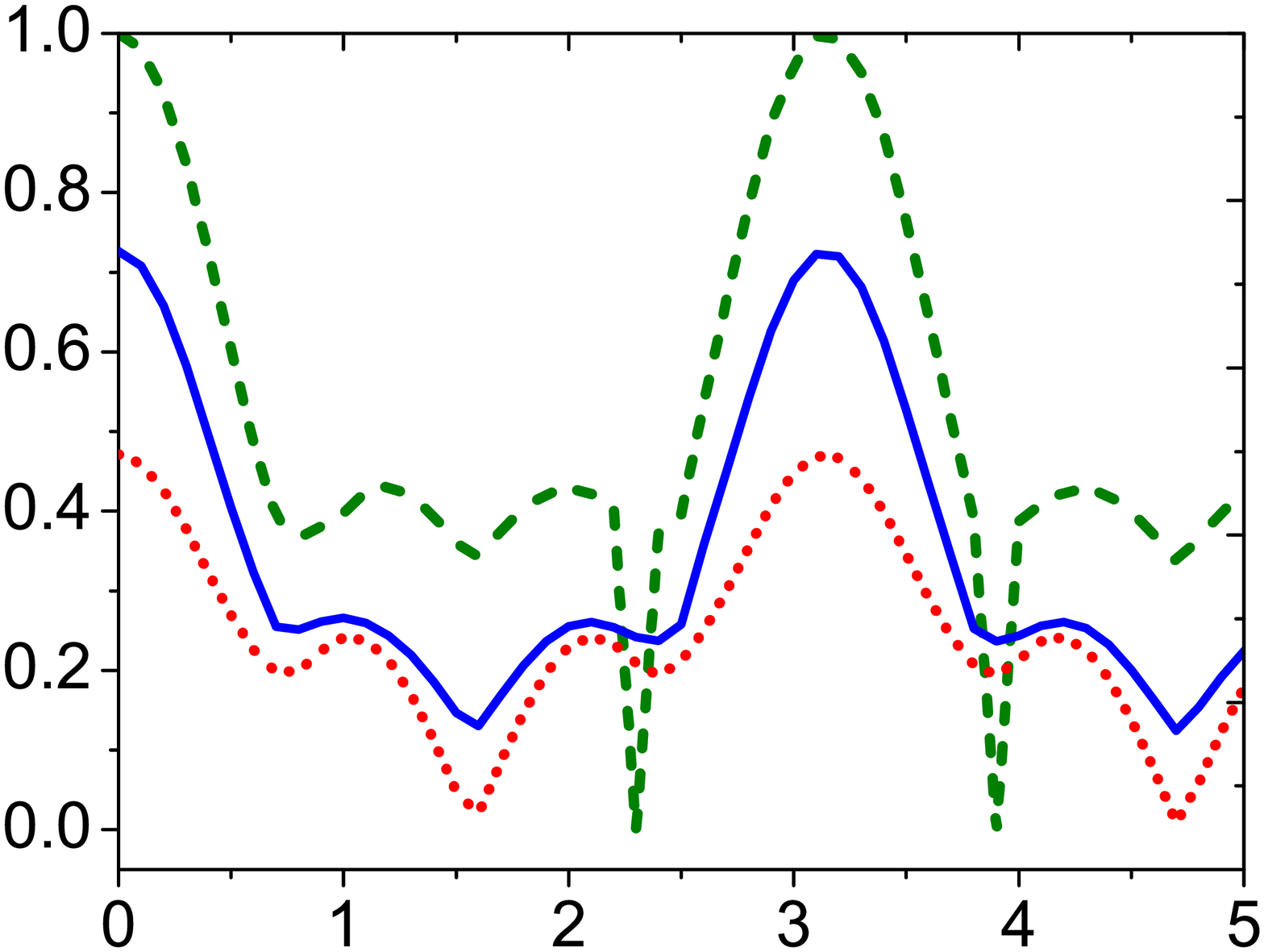}
 \put(-145,5){$\tau'$}
     \put(-275,90){$\mathcal{N}$}
      \caption{Entanglement of  two qutrits-system. The dash curve  for MES $(a_1=a_2=a_3=\frac{1}{\sqrt{3}}$)
      the dot  for  $(a_1=0,a_2=\frac{1}{\sqrt{3}}, a_3=\sqrt{\frac{2}{3}})$ and solid curves for
         $(a_1=0.3, a_2=0.4)$,
         respectively.}
\end{figure}

In Fig.(5), we plot the behavior of the entanglement when the
first qutrit is driven by the CERF. Let us consider that,  the system is
initially prepared in MES, i.e, we set $a_i=\frac{1}{\sqrt{3}},
i=1,2,3$. In this case, the general behavior shows that, the
entanglement is maximum at $\tau'=0$ (before the interaction is
switched on). As soon as the interaction between the first qutrit
and the classical external random field is switch on, the
entanglement decays as the interaction time is increased.
However, the entanglement vanishes for the first time
 at $\tau'\simeq 2.4$. For larger time, the
entanglement re-increases  gradually to reach its maximum value
($\mathcal{N}=1)$ for the first time  at $\tau'\simeq 3$.  This behavior is repeated while  increasing the interaction time between the qutrit and CERF.

 The effect of CERF on   qutrit-qutrit  system ( initially prepared in PES) is described also in Fig.(5), where two classes are
considered: the first  one is obtained by setting $a_1=0.3,
a_2=0.4$, while the
 second is obtained by setting $a_1=0, a_2=1/\sqrt{3}$. However,
 the phenomenon of entanglement sudden changes  is depicted for
 the two cases. For  less initial  entangled system, the lower
bounds of  entanglement are larger  than those which start with
larger degree of entanglement. Moreover, the upper  bounds of
entanglement are reached at the same time.

From the  behavior of  negativity, one can conclude that: as soon
as the interaction is switched on, the phenomenon of sudden decay
of entanglement is depicted  for MES and PES. The entanglement
suddenly increases to reach their maximum bounds which are
dependent  on the initial entanglement. However,  the upper and
lower bounds of entanglement are reached at the same time for all considered
states.

\section{Conclusion}
In this paper, the entanglement of three types of
different dimensional systems are discussed. It is assumed that only one subsystem is driven   by  an external random  field.
The suggested systems are: qubit-qubit, qubit-qutrit and qutrit
qutrit systems, which are described by $2\times 2,~ 2\times 3$ and
$3\times 3$ dimensions, respectively.  The general behavior shows
that, the entanglement is very sensitive to the classical external
random field. However, this sensitivity  depends on the size of the
driven subsystem and the initial degree of entanglement of the
system.

Maximum entangled states are the most sensitive states for the external field, where the phenomena of the  sudden decay of entanglement, changes, death and sudden re-birth are displayed  clearly.
On the other hand,  partially entangled states, are  found more robust
than the maximum entangled states to this classical external
random field.

The dimensions of the driven  subsystems by the external field
play an important role on the behavior of entanglement. This
result  can be observed clearly for the qubit-qutrit systems,
where if only the qubit is driven, then the sudden changes of entanglement ( death or birth) are faster than that shown  for the case in which  only the qutrit is driven by this external field. For the two parameters class, one can observe the phenomenon  of long-lived entanglement if the qubit is driven by the external field. However, for one parameter family, partially entangled states are more robust than
the two parameters class if the qutrit is driven by the classical
external random field.

For qutrit-qutrit system, the entanglement is more robust to the
classical external random field  than that displayed for qubit-qutrit systems, where the entanglement decays smoothly whereas  only the MES lose their entanglement for a very short period of  time and  the re-birth occurs again shortly. The effect of the dimensions of the driven
subsystems by the CERF can be observed by comparing the behavior
of  survival entanglement between the qubit-qubit and  the
qutrit-qutrit systems. It is clear that, the entanglement of
larger dimensional systems are more robust than smaller systems.

From  previous discussions one can conclude that: (i) maximum
entangled states are very sensitive to the classical external
random field; (ii) the entanglement and hence the information are
lost fast for smaller dimensional systems; (iii) one parameter
family is more powerful than the two parameters family, where the
possibility of losing its correlation is much smaller than that
shown for the two parameters family, and consequently their
efficiency to perform quantum information tasks is better; (iv)
the long-lived entanglement can be depicted for systems are
initially prepared in the two parameter family's class; (v) the
two-qutrit system is more robust than  the two-qubit system for
the classical external random field. Therefore we  can induce,
that in the context of quantum information and communication, the
2-qutrit systems are much better than the 2-qubit system in the
presence of classical external random field.

{\bf Acknowledgement:} We would like to thank the referee for
his/her comments which improve the manuscript.


\bigskip
\begin{thebibliography}{1}
\bibitem{Ban}
M. Ban,  S. Kitajima  and F. Shibata, "
Decoherence of entanglement in the Bloch channel"
J. Phys. A {\bf 38}, 4235
(2005); M. Ban, S. Kitajima and F. Shibata,"
Decoherence of quantum information in the non-Markovian qubit channel" J. Phys. A {\bf 38},
7161 (2005).
\bibitem{Eberly2006}
T. Yu, J.  Eberly: Opt. Commu "Sudden Death of Entanglement:
Classical Noise Effects" {\bf 264}, 393 (2006).

\bibitem{Metwally2009}N. Metwally,"Abrupt decay of entanglement and quantum communication through noise channels",
Quantum Inf Process, {\bf 9}  429 (2010).
\bibitem{Jia2014} J.-Dong Shi, T. Wu, X.Ke Song and Liu Ye,"Multipartite concurrence for X states under decoherence",
 Quantum
Inf. Process {\bf 13} 1045 (2014).
\bibitem{Aguilar2012}
O. J. Farías,  G. H. Aguilar, A. Valdés-Hernández,  P. H. Ribeiro, L.  Davidovich, S. P. Walborn,"Observation of the emergence of multipartite entanglement between a bipartite system and its environment", Phys Rev Lett. {\bf 109} 15:150403 (2012).



\bibitem{Aguilar2014}G. H. Aguilar, O. Jiménez Farías, A. Valdés-Hernández, P. H. Souto Ribeiro, L. Davidovich, and S. P. Walborn" Flow of quantum correlations from a two-qubit system to its environment", Phys. Rev. A {\bf 89}, 022339 (2014).


\bibitem{Agata2007}A. Checi$\grave{n}$ska and
K. W$\grave{o}$dkiewicz,"Separability of entangled qutrits in noisy channels", Phys. Rev. A {\bf 76} 052306 (2007).

\bibitem{Karpat2011} G. Karpat and Z. Gedik, "Correlation Dynamics of Qubit-Qutrit Systems in a Classical Dephasing Environment", Phys. Lett. A{\bf 375} 4166 (2011).

\bibitem{Liang2013} J.-Liang Guo, H. Li and G.-Lu Long," Decoherent dynamics of quantum correlations in qubit-qutrit systems", Quantum
Inf. Process {\bf 12} 3421-3435 (2013).

\bibitem{Khan} S. Khan and M. K. Khan, J. Mod. Opt. {\bf 58} 818 (2011).

\bibitem{Xiao} X. Xiao,"Protecting qubit–qutrit entanglement from amplitude damping decoherence via weak measurement and reversal", Phys. Scr.{\bf 88} 065102 (2014).

\bibitem{Franco0}A. D'Arrigo, R. Lo Franco, G. Benenti, E. Paladino, G. Falci,"Recovering Entanglement by Local Operations", Ann. Phys. {\bf 350}, 211–224
(2014).

\bibitem{Franco01}
Rosario Lo Franco, Bruno Bellomo, Erika Andersson, Giuseppe Compagno,"Revival of quantum correlations without system-environment back-action", Phys. Rev. A{\bf 85}, 032318 (2012)
\bibitem{Franco00} J.-S. Xu, K. Sun, C.-F. Li, X.-Y. Xu, G.-C. Guo, E. Andersson, R. Lo Franco and G. Compagno,"Experimental recovery of quantum correlations in absence of system-environment back-action",
Nature Commun. {\bf 4}, 2851 (2013).

\bibitem{Franco2}
A. Orieux, G. Ferranti, A. D'Arrigo, R. Lo Franco, G. Benenti, E.
Paladino, G. Falci, F. Sciarrino, and P. Mataloni,"Cavity-based
architecture to preserve quantum coherence and entanglement",
Sci.Rep.{\bf 5}, 8575 (2015).


\bibitem{Metwally2012} N. Metwally and S. S. Hassan," ," Information transfer and Orthogonality   speed via Pulsed- driven Qubit, ", J. Nonlinear Optics and quantum optics {\bf 60} 1-13 (2012).

\bibitem{Blanter2012} M. Dukalski and Ya M. Blanter,"Tripartite entanglement dynamics in a system of strongly driven qubits", J. Phys. B:At, Mol Opt. Phys. {\bf 45} 245504 (2012).

\bibitem{Metwally2014-1}
N. Metwally, H. A. Batarfi and S. S. Hassan,,"Long-lived entanglement with Pulsed-driven initially entangled qubit pai", Int. J. Quantum Information {\bf 12} 1450003 (2014).

\bibitem{Ying} Y. Jie Zhang, W. HAn, Y.-Je Xia, J. Peng Cao and H. Fan,"Classical-driving-assisted quantum speed-up", Phys. Rev A {\bf 91} 032112 (2015).

\bibitem{Obada0} A.-S Obada, A Eied and G M Abd Al-Kader,"Entanglement of a general formalism $\Lambda$-type three-level atom interacting with a non-correlated two-mode cavity field in the presence of nonlinearities", J. Phys. B: At. Mol. Opt. Phys. {\bf 41}
195503(2008).


\bibitem{Marlan}M. O. Scully and Shi-Yao Zhu,"Degenerate quantum-beat laser: Lasing without inversion and inversion without lasing", Phys. Rev. Lett {\bf 62} 2813
(1989).
\bibitem{Aguilar}
O. Aguilar, A. B. Klimov and H. de Guise,"Tomography vs quantum control for a three-level atom",
 Phys. Lett. A{\bf 359} 373-380 (2006).
\bibitem{Obada1} A.-S. F. Obada, A.A. Eied,"Entanglement in a system of an
E-type three-level atom interacting with a non-correlated two-mode cavity field in the presence of nonlinearities",
 Opt. Commun., {\bf 282} 2184-2191 (2009).




\bibitem{Obada2} M. Youssef, A.-S. F. Obada, N. Metwally," Some entanglement features of a three-atom Tavis–Cummings model: a cooperative case", J. Phys. B: At. Mol. Opt. Phys. 43 095501 (2010).

\bibitem{Obada3} A.-S. Obada, S. Hamoura and A. Eied,"Collapse-revival phenomenon for different configurations of a three-level atom interacting with a field via multi-photon process and nonlinearities", Eur. Phys. J. D {\bf 68} 18 (2014).

\bibitem{Lee} S. Lee, D. Chi, S. D. Oh, and J. Kim,"Convex-roof extended negativity as an entanglement measure for bipartite quantum systems", Phys. Rev. A {\bf 68} 062304 (2003).

\bibitem{Karpat}K. Ann, G. Jaeger," Entanglement sudden death in qubit-qutrit systems", Phys. Lett A{372} 579-583(2008); G. Karpat and Z. Gedik,"Correlation dynamics of qubit–qutrit systems in a classical dephasing environment", Phys. Lett. A {\bf 375} 4166 (2011).

\bibitem{Xing} Xing Xiao,"Protecting qubit-qutrit entanglement from amplitude damping decoherence via weak measurement and reversal", Phys. Scr. {\bf 88} 065102 (2014).
\bibitem{Metwally2014} N. Metwally,"Single and double changes of entanglement" J. Opt. Soc. Am B {\bf 31} 691
(2014).
\bibitem{Chi} D. Pyo Chi and S. Lee,"Entanglement for a two-parameter class of states in $2\otimes n$ quantum system", J. Phys. A: Math. Gen {\bf 36} 11503-11510 (2003).


\end{thebibliography}
\end{document}